\documentclass[11pt,a4paper,aps,nofootinbib]{article}
\usepackage{jcappub}
\usepackage[utf8]{inputenc}
\usepackage{color}
\usepackage{amsmath, bm, physics}
\usepackage{amssymb, mathtools}
\usepackage{subcaption}
\usepackage{hyperref}

\title{Probing the galactic and extragalactic gravitational wave backgrounds with space-based interferometers}

\author[a,1]{G. Mentasti,\note{Corresponding author.}}
\author[a]{C.~R. Contaldi,}
\author[b,c]{and M. Peloso}

\affiliation[a]{Blackett Laboratory, Imperial College London, SW7 2AZ, United Kingdom}
\affiliation[b]{Dipartimento di Fisica e Astronomia ``G. Galilei", Universit\`a degli Studi di Padova, via Marzolo 8, I-35131 Padova, Italy}
\affiliation[c]{INFN, Sezione di Padova, via Marzolo 8, I-35131 Padova, Italy}

\emailAdd{g.mentasti21@imperial.ac.uk}

\abstract{
We employ the formalism developed in \cite{Mentasti:2023gmg} and \cite{Bartolo_2022} to study the prospect of detecting an anisotropic Stochastic Gravitational Wave Background (SGWB) with the Laser Interferometer Space Antenna (LISA) alone, and combined with the proposed space-based interferometer Taiji.
Previous analyses have been performed in the frequency domain only. Here, we study the detectability of the individual coefficients of the expansion of the SGWB in spherical harmonics, by taking into account the specific motion of the satellites. This requires the use of time-dependent response functions, which we include in our analysis to obtain an optimal estimate of the anisotropic signal. We focus on two applications.
Firstly, the reconstruction of the anisotropic galactic signal without assuming any prior knowledge of its spatial distribution. We find that both LISA and LISA with Taiji cannot put tight constraints on the harmonic coefficients for realistic models of the galactic SGWB. We then focus on the discrimination between a galactic signal of known morphology but unknown overall amplitude and an isotropic extragalactic SGWB component of astrophysical origin. In this case, we find that the two surveys can confirm, at a confidence level $\gtrsim 3\sigma$, the existence of both the galactic and extragalactic background if both have amplitudes as predicted in standard models. We also find that, in the LISA-only case, the analysis in the frequency domain (under the assumption of a time average of data taken homogeneously across the year) provides a nearly identical determination of the two amplitudes as compared to the optimal analysis. 
}

\begin{document}
	
\maketitle
	
\section{Introduction}\label{sec:introduction}
The first direct observation of Gravitational Waves (GWs), generated by the merging of compact objects in 2015 was achieved by the LIGO-Virgo collaboration \cite{PhysRevLett.116.061102}. The observation gave rise to one of the most rapidly expanding fields
in contemporary physics. The technology implemented by the collaborations using interferometers for gravitational wave detection (currently operational ones include LIGO \cite{LIGOScientific:2014pky}, Virgo \cite{Acernese_2015},  and KAGRA \cite{10.1093/ptep/ptaa125}) is continuously improving. The increase in sensitivity and integration time is leading to improving upper limits on the amplitude of several models of the SGWB \cite{PhysRevD.104.022004,PhysRevD.105.122001}, with a detection of an astrophysical SGWB  being a realistic possibility over the next few years. 

Among the experiments that will collect GW data in the next decade, the space-based LISA mission~\cite{2019BAAS...51g..77T} is undoubtedly one of the most promising. Consisting of three satellites orbiting the Sun in elliptical orbits while maintaining a (nearly) constant distance of 2.5 million km from each other, the instrument will be capable of observing gravitational waves in a frequency range that is inaccessible to ground-based interferometers. LISA is currently in the definition phase, where the mission system requirements are being reviewed. This phase will end with possible adoption in 2024, which would see the experiment fly in $\sim$2035 and begin data collection a few years later \cite{CERN_talk,2019BAAS...51g..77T}.

On the other hand, Taiji \cite{doi:10.1142/S0217751X2050075X,Liang:2021bde}, an analogous experiment proposed by the Chinese Academy of Sciences and currently in the planning stages, may be operative by the time that LISA will be flying. The proposed Taiji experiment has slightly longer arms than LISA (3 million km) and a sensitivity that peaks in the same frequency range.
	
At these frequencies, we expect many physical phenomena that can produce a GW signal to be active. In this work, we focus on the class of sources that cannot be resolved as individual events, but which can produce a globally detectable effect given their large number. The signal coming from the superposition of all the individually undetectable sources of GW radiation is called the Stochastic Gravitational Wave Background (SGWB).

While only upper bounds exist at frequencies probed by ground-based interferometers \cite{PhysRevD.104.022004,LIGOanisotropic2022}, 
interest in SGWBs has increased significantly over the past year after the recent claims by Pulsar Timing Array (PTA) collaborations \cite{EPTA:2023fyk,NANOGrav:2023gor,refId0} of a detection of a GW signal at much lower frequencies than those probed by LISA. These measurements may be evidence of an SGWB at frequencies $\sim 10^{-9} - 10^{-8}$ Hz.
Analogously, we expect a SGWB to exist at LISA and Taiji frequencies. These would be sourced from several different phenomena. On the one hand, there is an astrophysical component of the SGWB, mainly constituted by the GW signal coming from mergers of compact binary systems, such as stellar mass black holes and neutron stars. On top of this signal, there are GWs generated by other astrophysical sources, such as massive black holes (that are expected from several models explaining the formation of Supermassive Black Holes (SMBH)), other galactic binaries, and Extreme Mass Ratio Inspirals (EMRIs).
\cite{Amaro-Seoane2023}. On the other hand, in the LISA frequency band, a cosmological GW background may also be present~\cite{LISAcosmogroup}. In this case, the mechanisms capable of sourcing a GW signal can have different origins. There is the possibility to observe a background that originates from inflation, preheating, phase transitions, topological defects, primordial black holes (see \cite{LISACosmologyWorkingGroup:2023njw} for a general overview), or other (known or as yet unknown) early universe phenomena.

The variety of phenomena mentioned above, together with the fact that we can only access their superposition in the background, results in many challenges from the point of view of data analysis. However, the distinct shapes in frequency and angular distribution of the Power Spectral Density (PSD) of the various signals can be exploited to better model the effect and to optimize the analysis process.
In particular, thanks to the directional dependence, we may be able to distinguish between galactic and extragalactic components and search for any correlation with known tracers of structure \cite{SGWB-astro-sources,SGWB-astro-correl,SGWB-astro-correl-1,Ricciardone:2021kel}.
This paper aims to exploit the data analysis techniques described in \cite{Mentasti:2023gmg} within the framework of ground-based detectors and to apply them to the LISA data analysis to produce the final forecast on the observability of some features of the SGWB. We focus on detecting the anisotropies of an SGWB~\cite{PhysRevD.106.023010,10.1093/mnras/stab2479}, taking advantage of the periodic motion of LISA, and then of LISA in combination with Taiji in their trajectory around the Sun, along with a complete time-frequency analysis.
A network of constellations of space-based interferometers has also been considered in \cite{Capurri_2023}; in contrast to that work, we leverage our analysis on the specific trajectory of the satellites, with the application to the LISA - Taiji pair. 

Further information on the origin of the SGWB can be obtained by studying its pola\-rization~\cite{Seto:2007tn,Crowder:2012ik,Smith:2016jqs,anisotropy-DomckePeloso,Orlando_2021,Martinovic:2021hzy}. A coupling between a pseudo-scalar inflaton and a gauge field, can result in a fully circularly polarized SGWB~\cite{Sorbo:2016rzu}. Still, a degree of net polarization might also appear in the astrophysical SGWB due to Poisson fluctuations in the (finite) number of unresolved sources~\cite{ValbusaDallArmi:2023ydl} and might be potentially detectable for a network of space based interferometers \cite{PhysRevD.107.104040}. 
As explained below, in our treatment we consider an intrinsically stationary background, i.e. whose statistical proprieties do not change with time. This assumption is reasonable for cosmological and astrophysical backgrounds made of a superposition of many unresolved sources, as predicted in most theoretical models. However, analysis can be extended to non-stationary cases \cite{popcorn,PhysRevD.107.103026}. 

The plan of the paper is as follows. In section~\ref{sec:theSGWB}, we summarize the most useful definitions in setting up the analysis of SGWBs and introduce the theoretical parameters we would like to probe. This is done following the formalism of~\cite{Mentasti:2023gmg} and~\cite{Bartolo_2022}. In section~\ref{sec:analysis} we go beyond what was developed in \cite{Mentasti:2023gmg} to analyze the case of LISA and Taiji, showing how one can build physical observables that can be used to probe the theoretical features of a SGWB. We show how to perform the optimal analysis, taking into account the nuisance effects due to instrumental noise and the variance of the signal as we did for ground-based detectors in \cite{Mentasti:2023gmg}, defining an optimally filtered chi-squared in the measurement.
We then produce the forecasts of two analyses that can be performed using this formalism. In section~\ref{sec:gal_map} we show how LISA alone and LISA-Taiji combined can constrain the anisotropic coefficients of the map of the astrophysical background of our galactic binaries \cite{10.1093/mnras/stac3686,PhysRevD.104.043019}, along with the total energy density.
 On the other hand, in section~\ref{sec:galaxy+back} we show how accurately the two networks of instruments can disentangle a galactic and an extragalactic astrophysical component of the SGWB \cite{Babak_2023}, taking advantage of the different frequency dependencies of the power spectra and the different angular distribution of the sources. In the last section~\ref{sec:conclusions} we summarize the results of our analysis and discuss the advantages and limitations of the two experiments in both applications.

\section{The stochastic gravitational wave background}
\label{sec:theSGWB}
The starting point of our analysis is the formalism that we developed in \cite{Mentasti:2020yyd,Mentasti:2023gmg} to determine the observability of anisotropic modes of SGWBs with generalized networks of baselines. We refer the interested readers to these works for details.
Generally, we consider gravitational waves at $x^\alpha$, the location of the detectors, as small perturbations of the metric tensor on a Minkowski background in such a way that it can be decomposed in plane waves (the usual $h_{00} = h_{0a} = h_{aa} = \partial_a h_{ab}   = 0$ gauge has been adopted, where Latin indices run only over the spatial components)
	\begin{align}
		h_{ab}(t,\vec x)&=\int_{-\infty}^{\infty}df\int d^2 {\hat n} \, e^{2\pi i f \left( t - {\hat n} \cdot \vec{x} \right)} 
		\sum_{\lambda=R,L} h_\lambda (f,\hat n) \, e^\lambda_{ab} \left( {\hat n} \right)\,,
		\label{eqGWTT}
	\end{align}
	where $\hat n$ is the unit vector on the $2$-sphere, and $e^R_{ab}(\hat n)$, $e^L_{ab}(\hat n)$ are the two chiral polarization basis tensors, with the same normalization adopted in \cite{Bartolo_2022}, i.e. $e^{\lambda*}_{ab}e^{\lambda'}_{ab}=\delta_{\lambda\lambda'}$. 
The direction vector $\hat n$ is defined as $\hat n=\cos\phi\sin\theta\,\hat x +\sin\phi\sin\theta\,\hat y +\cos\theta\,\hat z$.

	We define \eqref{eqGWTT} in ecliptic reference coordinates, setting the $\hat{z}$ axis to point along the direction perpendicular to the plane of the Earth orbit.
	To describe a stochastic source, we treat the complex amplitude $h_\lambda(f,\hat n)$ as a random Gaussian variable with zero mean. This means that its statistics are completely specified by its variance. The dependence of the stochastic background on frequency and direction may be stated solely in terms of the expectation value of the two-point correlator for the random variable $h_\lambda(f,\hat n)$ as
	\begin{align}
		\left\langle h^*_\lambda \left( f, {\hat n} \right) h_{\lambda'} \left( f', {\hat n}' \right) \right\rangle = \delta_{\lambda\lambda'} \delta_D^{(2)} \left( {\hat n} - {\hat n}' \right) \delta_D \left( f-f' \right) \mathcal{H}(|f|,\hat n)\,,
		\label{eq2points}
	\end{align}
	where $\delta_D^{(2)} \left( {\hat n} - {\hat n}' \right)$ is a covariant two-dimensional Dirac delta-function on the unit two-sphere and $\delta_D \left( f-f' \right)$ the Dirac delta function on the frequency space. 
	In some cases, we suppose that we have a factorized dependence of the angles and the frequencies in the power spectrum, such as
	\begin{align}
		\mathcal{H}(|f|,\hat n)&= \frac{\tilde I_{00}\left( \left\vert f \right\vert \right)}{4\pi}  P \left( {\hat n} \right)\,,
	\end{align}
	where the function $\tilde I_{00}\left( \left\vert f \right\vert \right)$ is introduced analogously in \cite{Bartolo_2022}.
Although commonly adopted in the literature, this factorization is not generic, and, in fact, in section \ref{sec:galaxy+back} we consider a more complex angular and frequency dependence.
It is conventional to define the fractional energy density in GW per logarithmic frequency interval and per polarisation as
	\begin{equation}
		\Omega_{GW} \left( f \right)\equiv \frac{f}{\rho_{\rm crit} }\frac{d\rho_{GW,\lambda}}{df} = \frac{4\pi^2f^3}{3H_0^2}\tilde I_{00} \left(|f| \right)  \,,
		\label{OmegaGW2}
	\end{equation}
	where $\rho_{\rm crit}$ is the current energy density of the universe, which we assume to be flat, while $H_0$ is the Hubble constant. In the second step, we have used the fact that only the monopole contributes to the total GW energy density (after integrating over the arrival direction). 
Next, we decompose the angular power spectrum in terms of spherical harmonics, writing it as
	\begin{equation}\label{eqPOmega}
		P(\hat n)=\sum_{\ell=0}^{+\infty}\sum_{m=-\ell}^{\ell} \, \delta^{GW}_{\ell m} \, Y_{lm}(\hat n) \;\;\;,\;\;\; \delta^{GW}_{00} = \sqrt{4 \, \pi} \,, 
	\end{equation}
	where the value of $\delta^{GW}_{00}$ is set considering the convention for the monopole $Y_{00} = \frac{1}{\sqrt{4 \pi}}$.
	Finally, we rewrite the frequency dependence of $\Omega_{\rm GW} \left( f \right)$ as
	\begin{equation}
		\Omega_{\rm GW} \left( f \right) = {\Omega}_{0} \,\psi \left( f \right) \;\;,\; {\rm with } \; \psi \left( f_0 \right) = 1 \;,
		\label{PowerLawOmega}
	\end{equation}
	so that ${\Omega}_{0}$ is the fractional energy density at the pivot frequency $f_0$. In order to produce numerical forecasts in our work, we suppose that we have a perfect knowledge of this function. However, it is in principle possible to write it in parametric form and infer the value of the parameters in the same analysis.
	
	\section{SGWB effect in space-based detectors}\label{sec:analysis}
	
\begin{figure}[ht!]
		\centerline{
			\includegraphics[width=0.8\textwidth,angle=0]{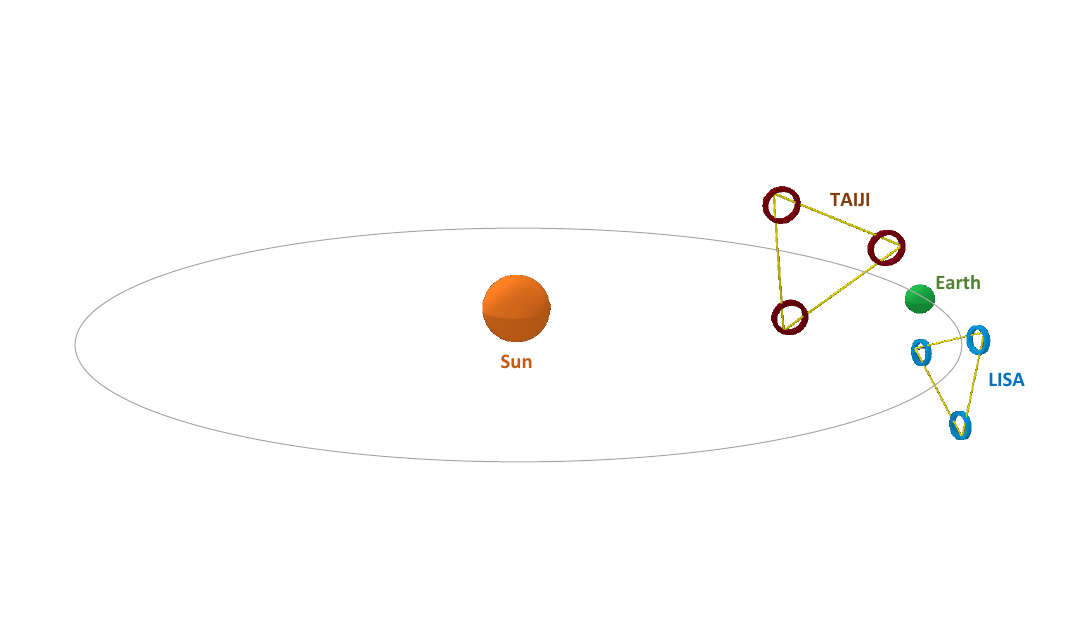}
		}
		\caption{Scheme of the \textit{LISA-Taiji} configuration. The three LISA satellites are separated by an arm length of $2.5\times 10^6$~km, while for Taiji the arm length is $3\times 10^6$~km. The center of mass of each of the two constellations lays at $20^\circ$ from the position of the Earth in the ecliptic plane, and it follows the trajectory of the Earth during its rotation around the Sun. The precise trajectory of the six satellites is described in appendix~\ref{app:positions}. We refer as \textit{LISA-self} configuration the same in the figure where however the three Taiji satellites are not present.}
		\label{fig:LISA-Taiji-scheme}
	\end{figure}

	We consider two possible configurations of space-based interferometers. We call \textit{LISA-self} the network made of three satellites in a triangular-shaped configuration that constitutes the LISA experiment as in figure
	\ref{fig:LISA-Taiji-scheme}.
	On the other hand, we call \textit{LISA-Taiji} the network made of the 3 LISA satellites along with Taiji, a proposed experiment by the Chinese Academy of Sciences \cite{doi:10.1142/S0217751X2050075X,Ruan2020} that would be built similarly to LISA, having 3 satellites in a triangular-shaped configuration as in figure \ref{fig:LISA-Taiji-scheme}. We denote by $\Delta T_{ij}^L(t)$ the modification to the time of flight of a photon in the laser beam between the i-th and the j-th satellite of LISA due to the effect of a gravitational wave, where $i,j=\{1,2,3\}$. Analogously, $\Delta T_{ij}^T(t)$ will be the same time of flight distortion between two satellites in the Taiji constellation. One can show that
	\begin{align}
		\Delta T_{ij}^{L}(t)=\frac{\hat l_{ij}^{L,a}(t)\hat l_{ij}^{L,b}(t)}{2}\int_0^{L^{L}}ds\,h_{ab}(t(s),\vec x(s))\,,
	\end{align}
	where $\hat l_{ij}^L(t)$ is the unit vector going from $\vec x_i^L$ to $\vec x_j^L$, the i-th and j-th satellite of LISA. An analogous expression holds for Taiji, with the suffix L replaced by the suffix T in all these quantities. 
 In our numerical evaluations, the unperturbed distance $L^L$ between two LISA satellites is set equal to $2.5\times 10^6$~km, while for Taiji the unperturbed distance $L^T$ is $3\times 10^6$~km \cite{Ruan2020}.  The actual trajectory considered for the six satellites is explained in full detail in appendix~\ref{app:positions}.
	
	Following the formalism adopted in \cite{Bartolo_2022}, we introduce the set of three Time-Delayed Interferometric (TDI) measurements $\Delta F_i^{L}(t)$ made by a combination of the Doppler frequency shifts between the LISA satellites ($\Delta F_i^{T}(t)$ will be the same combinations for the Taiji constellation). We reference (4.11) in \cite{Bartolo_2022} for the precise definition of these quantities. Under the assumption of equal arm lengths (which means that both LISA and Taiji are perfect equilateral triangles; in appendix~\ref{app:positions} we verify the accuracy of this assumption), we can combine the TDI data streams $\Delta F_i^{L/T}(t)$ in order to diagonalize the noise matrix. This accounts for performing the usual linear transformation in the "AET" basis
	\begin{align}
		\Delta F_O^{L/T}(f,t)&=c_{Oi}\int_{t-\tau/2}^{t+\tau/2} dt'\,e^{-2\pi i ft'}\Delta F_i^{L/T}(t')\,,\qquad O\in\{A,E,T\}\,,\nonumber\\
		c &\equiv  \left( \begin{array}{ccc} 
			-\frac{1}{\sqrt{2}}  & 0 & \frac{1}{\sqrt{2}} \\ 
			\frac{1}{\sqrt{6}} & - \frac{2}{\sqrt{6}} & \frac{1}{\sqrt{6}} \\ 
			\frac{1}{\sqrt{3}} &\frac{1}{\sqrt{3}} & \frac{1}{\sqrt{3}}
		\end{array} \right) \;.   
		\label{def-cmat}
	\end{align}
	After some straightforward algebra (see \cite{Bartolo_2022} and \cite{Flauger_2021} for more details) the self-correlators between each pair of the LISA TDI data streams in the "AET" basis read 
	\begin{align}
		&\langle \Delta F_O^L(f,t)\Delta F_{O'}^{L,*}(f,t)\rangle = 2\tau\left|\frac{f}{f_*^{L}}W^L(f)\right|^2\tilde I_{00}(f)\frac{1}{8\pi}\int d^2\hat n \sum_{\ell m} \delta_{\ell m}^{GW}Y_{\ell m}(\hat n)\times\nonumber\\
		&\times\sum_{i,j}c_{Oi}c_{O'j}\sum_\lambda R_{\lambda}(f\hat n,\hat{l}_{i,i+1}^L,\hat{l}_{i,i+2}^L)R_{\lambda}^{*}(f\hat n,\hat{l}_{j,j+1}^L,\hat{l}_{j,j+2}^L)\,e^{-2\pi i f \hat n\cdot (\vec x_i^L-\vec x_j^L)}\,,
	\end{align}
	with $f_*^{L/T}=\frac{1}{2\pi L^{L/T}}$. On the other hand, the cross-correlators between the "AET" channels of LISA and the "AET" channels of Taiji are
	\begin{align}
		&\langle \Delta F_O^L(f,t)\Delta F_{O'}^{T,*}(f,t)\rangle = 2\tau\frac{f^2}{f_*^{L}f_*^{T}}W^L(f)W^{T,*}(f)e^{2\pi i f(L^T-L^L)}\tilde I_{00}(f)\frac{1}{8\pi}\int d^2\hat n \sum_{\ell m} \delta_{\ell m}^{GW}Y_{\ell m}(\hat n)\times\nonumber\\
		&\times\sum_{i,j}c_{Oi}c_{O'j}\sum_\lambda R_{\lambda}(f\hat n,\hat{l}_{i,i+1}^L,\hat{l}_{i,i+2}^L)R_{\lambda}^{*}(f\hat n,\hat{l}_{j,j+1}^T,\hat{l}_{j,j+2}^T)\,e^{-2\pi i f \hat n\cdot (\vec x_i^L-\vec x_j^L)}\,,
	\end{align}
	where the time dependence of $\vec x_i^{L/T}(t)$ and $\hat l_{ij}^{L/T}(t)$ is understood and where the explicit expression of the functions $W^{L/T}(f)$ and $R_\lambda (f\hat n,\hat l_{ij},\hat l_{ik})$ can be found in eqs.~(4.14) and~(A.4) of \cite{Bartolo_2022}, respectively.
	As stated above, the actual measurement in the detector can be affected by some instrumental noise in such a way that 
	\begin{align}
		\tilde m_O^{L/T}(f,t)=\Delta F^{L/T}_O(f,t)+\tilde n^{L/T}_O(f,t) \;, 
	\end{align}
	where we assume that the noise $\tilde n^{L/T}_O(f,t)$ is drawn from a Gaussian distribution and that noises of different detectors are uncorrelated, i.e.
	\begin{align}
		\langle n_O^{L/T}(f)n_{O'}^{L/T}(f')\rangle&=\frac 1 2 \delta(f-f')\delta_{OO'}N^{L/T}_{O}(f) \;, \nonumber\\
		\langle n_O^{L}(f)n_{O'}^{T}(f')\rangle&=0 \;. 
	\end{align}
	The functions $N^{L/T}_{O}(f)$ are given in eqs. (B14) and  (B15) in \cite{Bartolo_2022}.~\footnote{For what concerns the two parameters of the noise PSD of LISA and Taiji, we set $P=15$ and $A=3$ in the LISA curve~\cite{Caprini_2019}. For what concerns Taiji, we choose $P=8$ and $A=3$~\cite{doi:10.1142/S0217751X2050075X,Ruan2020}. 
 }
	We can now build the final estimator from the self-correlators of LISA TDI datastreams:
	\begin{align}\label{CLL}
		\mathcal{C}^{LL} \equiv \sum_{O, O^{\prime}} \int_0^T d t \int_{-\infty}^{+\infty} d f\left[\tilde{m}^L_O(f, t) \tilde{m}^{L,*}_{O^{\prime}}(f, t)-\left\langle\tilde{n}^L_O(f, t) \tilde{n}^{L,*}_{O^{\prime}}(f, t)\right\rangle\right] \tilde{Q}_{O O^{\prime}}(t, f) \;,
	\end{align}
	where $\tilde{Q}^{LL}_{O O^{\prime}}(t, f)$ are filter functions to be optimized to maximize the signal-to-noise ratio of the measurement.
	In a similar way, the estimator built by making use of the cross-correlators between LISA and Taiji datastreams is
	\begin{align}\label{CLT}
		\mathcal{C}^{LT} \equiv \sum_{O, O^{\prime}} \int_0^T d t \int_{-\infty}^{+\infty} d f\tilde{m}^L_O(f, t) \tilde{m}^{T,*}_{O^{\prime}}(f, t)\tilde{Q}^{LT}_{O O^{\prime}}(t, f)\,.
	\end{align}
We stress a fundamental difference between the two analyses considered. In the first case (when only LISA data streams are available) a perfect knowledge of the noise curve is assumed: any error in the estimate of the noise will translate into a bias in the final estimator $\mathcal{C}^{LL}$ \cite{Muratore:2023gxh}. On the other hand, since it is reasonable to assume no correlation between LISA and Taiji realizations of the noise, in the second analysis an imprecise estimate of the noise within each constellation would not bias the expectation value of the estimator $\mathcal{C}^{LT}$. This is a well-known fact in analyses with ground-based detectors, where self-correlators of measurements of the same detector are typically not included in the analysis to avoid biases from the imperfect knowledge of the noise function $N$ \cite{Allen:1996gp}. For this reason, in the LISA-Taiji analyses we disregard any self-correlators of two measurements taken within one constellation.
	
	\subsection{Signal expectation value and chi-squared}
	After the same steps performed in \cite{Bartolo_2022}, we find that the expectation value of the estimator \eqref{CLT} is
	\begin{align}
		\langle \mathcal{C}^{LT}\rangle&=\sum_{OO'}\frac{\tau}{2}\int_0^T dt\int_0^{+\infty}df\frac{\tilde I_{00}(f)}{\sqrt{4\pi}}W^L(f)W^{T,*}(f)e^{2\pi i f(L^T-L^L)}\times\nonumber\\
		&\times\sum_{\ell m}\delta_{\ell m}^{GW}\tilde R^{\ell m}_{OO'}(f,t)\left[\tilde Q_{OO'}(t,f)+\tilde Q_{O'O}(t,-f)\right]\,,
	\end{align}
	where
	\begin{align}\label{R_oop_lm}
		\tilde R^{\ell m}_{OO'}(f,t)&=\sum_{i,j}c_{Oi}c_{O'j}\frac{1}{2\sqrt{4\pi}}\int d^2\hat n\,e^{-2\pi i f \hat n\cdot (\vec x_i^L-\vec x_j^L)}Y_{\ell m}(\hat n)\times\nonumber\\
		&\times\sum_\lambda R_{\lambda}(f\hat n,\hat{l}_{i,i+1}^L,\hat{l}_{i,i+2}^L)R_{\lambda}^{*}(f\hat n,\hat{l}_{j,j+1}^T,\hat{l}_{j,j+2}^T)\,.
	\end{align}
	An analogous formula holds for $\langle\mathcal{C}^{LL}\rangle$ and can be found as  (4.36) in \cite{Bartolo_2022}.
	Next, as done in \cite{Mentasti:2023gmg}, we build a chi-squared in the quantity
	\begin{align}
		r^{RL}(\theta)\equiv \mathcal{C}^{LT}(\theta)-\langle \mathcal{C}^{LT}(\hat\theta)\rangle\,,
	\end{align}
	where $\theta=(\Omega_0,\{\delta_{\ell m}^{GW}\})$ are the parameters of the theory that we want to probe and $\hat\theta$ their fiducial values.
	Following the analogous steps in \cite{Mentasti:2023gmg}, we obtain the final expression of the optimal filtered chi-squared on the estimator for the LISA-Taiji analysis
	\begin{align}\label{chi_sq_cross_general}
		\chi^2_{LT}&=\left(\frac{3H_0^2}{4\pi^2\sqrt{4\pi}}\right)^2 \sum_{OO'}\int_{0}^{\infty} df \frac{\hat\Omega_0^2}{f^6}\psi^2(f)\int_0^Tdt\frac{1}{D_{OO'}(f,t)}\left|\sum_{\ell m}\left(\frac{\Omega_0}{\hat \Omega_0}\delta^{GW}_{\ell m}-\hat\delta^{GW}_{\ell m}\right)\tilde R^{\ell m}_{OO'}(f,t)\right|^2 \;,
  \end{align}
where 
\begin{align}
  \nonumber\\
		D_{OO'}(f,t)&\equiv\mathcal{T}^*_{OO}(f,t)\mathcal{T}_{O'O'}(f,t)+\mathcal{T}^*_{OO'}(f,t)\mathcal{T}_{O'O}(f,t)\,,\nonumber\\
\label{DOOp}
\end{align}
and where
\begin{align} 
  \mathcal{T}_{OO'}(f,t)&\equiv\sum_{\ell m}\tilde{I}_{\ell m}(f)\tilde R^{\ell m}_{OO'}(f,t)+\delta_{OO'}\tilde N_O(f)=\nonumber\\
		&=\hat \Omega_0\,\psi(f)\frac{3H_0^2}{4\pi^2\sqrt{4\pi}f^3}\sum_{\ell m}\delta_{\ell m}^{GW}\tilde R^{\ell m}_{OO'}(f,t)+\delta_{OO'}\tilde N_O(f)\,.
	\end{align}
We note that $D_{OO'}(f,t)$ is a real quantity. This follows from the fact that, assuming a real map for the anisotropic distribution of the galactic signal (as  \eqref{map_Galaxy} is), we have $\mathcal{T}_{O'O}(f,t)=\mathcal{T}_{OO'}^*(f,t)$.
In these expressions, the index $O$ runs over the LISA channels and the index $O'$ over the Taiji ones. The chi-squared evaluates to 
	\begin{align}\label{chi-squared-cross_begin}
		\chi^2_{LT}=\left(\frac{3H_0^2}{4\pi^2\sqrt{4\pi}}\right)^2 \sum_{OO'}\int_{0}^{\infty} df \frac{\hat\Omega_0^2\,\psi^2(f)}{f^6}\int_0^Tdt\frac{1}{D_{OO'}(f,t)}\left|\sum_{\ell m}\left(\frac{\Omega_0}{\hat \Omega_0}\delta^{GW}_{\ell m}-\hat\delta^{GW}_{\ell m}\right)\tilde R^{\ell m}_{OO'}(f,t)\right|^2\,,
	\end{align}
	and we note that, as in the analysis of~\cite{Bartolo_2022}, the TDI factors $W^{L,T}(f)$ cancel out from this expression.~\footnote{More precisely, compare this expression with  (4.38) of ~\cite{Bartolo_2022}. Using a schematic notation, the $\chi^2$ in that equation is proportional to the ratio between $R = W^2 {\tilde R}$ and $N \propto W^2 {\tilde N}$, and the TDI factors $W$ cancel in this ratio, leading to an expression analogous to (\ref{chi-squared-cross_begin}).}
	\begin{figure}[h!]
		\centerline{
			\includegraphics[width=0.8\textwidth,angle=0]{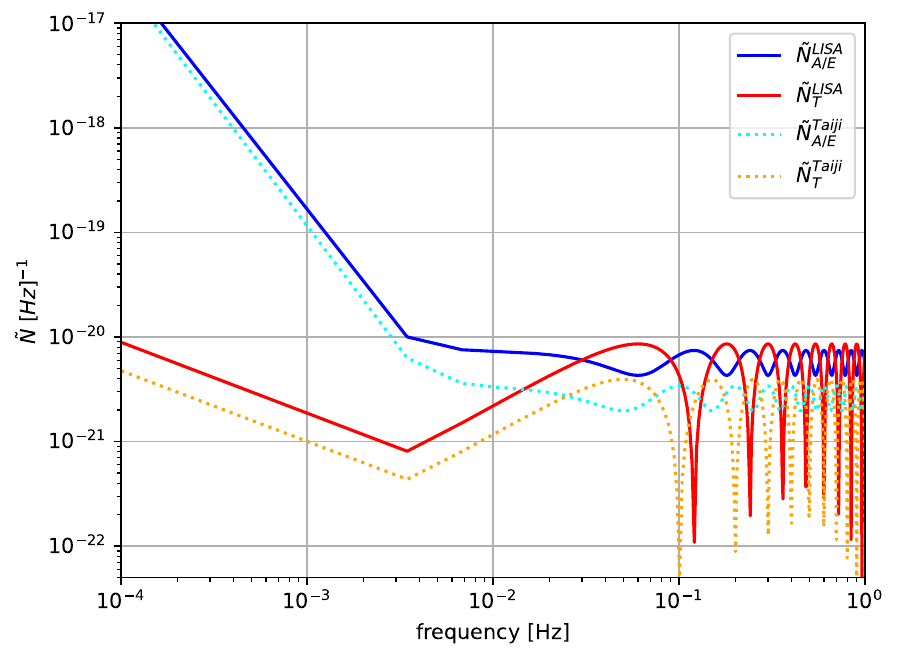}
		}
		\caption{Freuquency dependence of the noise PSD $\tilde N_O^{L/T}(f)$ as defined in eqs. (B14) and (B15) in \cite{Bartolo_2022}. As discussed in the text, for LISA we have chosen a value for the parameters $P=15$, $A=3$, and $L=2.5\times 10^6$~km, while the PSD of Taiji is computed assuming $P=8$, $A=3$, and $L=3\times 10^6$~km. }
\label{fig:allpsd}
	\end{figure}
	On the other hand, for what concerns the LISA-self analysis, the corresponding equation of \eqref{chi-squared-cross_begin} is  (4.38) in \cite{Bartolo_2022}.
	We perform the integration in time of \eqref{chi-squared-cross_begin} numerically, by taking advantage of the fact that the response functions $R^{\ell m}_{OO'}(f,t)$ are periodic in time with period $T_s=1$~yr. Therefore we replace the upper time of integration in ~(\ref{chi-squared-cross_begin}) with $T_s$ and we multiply the results by the number of years of observation $n$. Furthermore, as we are unable to perform the time integration analytically, we discretize it in a large but finite amount $N_t$ of time intervals. The numerical evaluation of the time-dependent response function is extremely expensive, and so we evaluate it once in each time interval, at the times $t_i=i\frac{T_s}{N_t}$ with $i\in\{1,\dots, N_t\}$. The quantity $N_t$ is limited from below by the requirement that the observed sky does not change appreciably during each time interval. More precisely, as we assume stationary signal statistics, we require that the constellations vary their angular position (in their motion around the sun) by an angle $\ll \frac{180^\circ}{l_{\rm max}}$, where $l_{\max}$ is the maximum multipole that we include in our analysis (choosing smaller values of $N_t$ would smear the measurement of the higher multipoles).
	For $l_{\rm max} = 30$, the limiting angle is $6^\circ$, which is spanned in about $6$ days. Correspondingly, we must choose $N_t \gg 60$. The quantity $N_t$ is also limited from above, by the requirement that all frequencies of our interest are well probed within each time window. The smallest frequencies to which the measurement is sensitive is $f_{\rm min} = {\rm O } \left( 10^{-4} \, {\rm Hz} \right)$. This translates into $\frac{T_s}{N_t} \, f_{\rm min} \gg 1$, which gives $N_t \ll 3000$. In our analysis, we choose $N_t = 365$. With the discretization in time, ~\eqref{chi_sq_cross_general} is cast in the form 
	\begin{align}\label{chi_sq_LT_discrete}
		\chi^2_{LT}&=n\,T_s\left(\frac{3H_0^2}{4\pi^2\sqrt{4\pi}}\right)^2 \sum_{OO'}\int_{0}^{\infty} df \frac{\hat\Omega_{0}^2\,\psi^2(f)}{f^{6}}\times\nonumber\\
		&\times\frac{1}{N_t}\sum_{i=1}^{N_t}\frac{1}{D_{OO'}(f,t_i)}\left|\sum_{\ell m}\left( \frac{\Omega_0}{\hat \Omega_0} \, \delta^{GW}_{\ell m}-\hat\delta^{GW}_{\ell m}\right)\tilde R^{\ell m}_{OO'}(f,t_i)\right|^2\,. 
	\end{align}
 In the following sections, we employ this procedure for two different analyses. Firstly, in section \ref{sec:gal_map} we forecast the measurement of the energy density $\Omega_0^2$ and of the anisotropic coefficients $\delta_{\ell m}^{GW}$ for a SGWB generated by the galactic sources only. As a second example, in section \ref{sec:galaxy+back} we consider a SGWB resulting from the superposition of a galactic and an extragalactic signal, forecasting the observability of the energy density of the two components.

\section{Reconstructing the galaxy map}\label{sec:gal_map}
	
	As a first example, in this section we would like to show how LISA and LISA with Taiji can reconstruct the anisotropic distribution of the galactic background, assuming that this is the only component of the SGWB. 
	To do so, we first assume that the frequency dependence of the power spectrum is~\cite{PhysRevD.104.043019}
	\begin{align}\label{psi_galaxy}
\psi(f)\equiv\psi_G(f)= A\,e^{-(f/f_1)^\alpha}\left[1+\tanh\left(\frac{f_{\rm knee}-f}{f_2}\right)\right] \;, 
	\end{align}
	which is the best fit for the SGWB after the removal of the resolvable sources.~\footnote{The numerical values of the constants $\alpha,f_1,f_{\rm knee},f_2$ are taken from \cite{10.1093/mnras/stac3686} and the overall constant A is set such that $\psi_G(f_0=1\text{mHz})=1$.} In agreement with \cite{10.1093/mnras/stac3686}, we have a fiducial value of $\hat\Omega_0=2\times 10^{-11}$. As a second assumption, the model of the galaxy used to set the fiducial value of the anisotropic coefficients $\hat\delta_{\ell m}^{GW}$ is explained in appendix~\ref{app:galactic_deltalm}.
	
\begin{figure}[h!]
\centering
\begin{subfigure}[b]{0.7\textwidth}
\includegraphics[width=\textwidth,trim=20 30 20 22, clip]{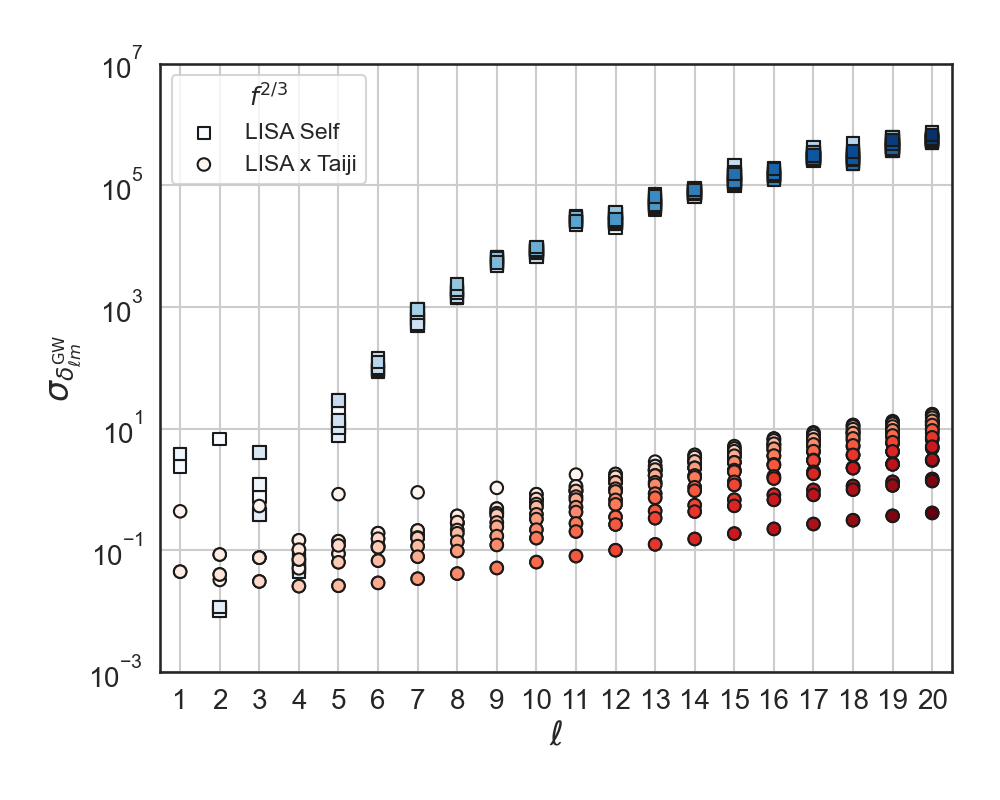}    
\end{subfigure}
\hfill
\begin{subfigure}[b]{0.7\textwidth}  
\centering 
\includegraphics[width=\textwidth,trim=20 30 20 22, clip]{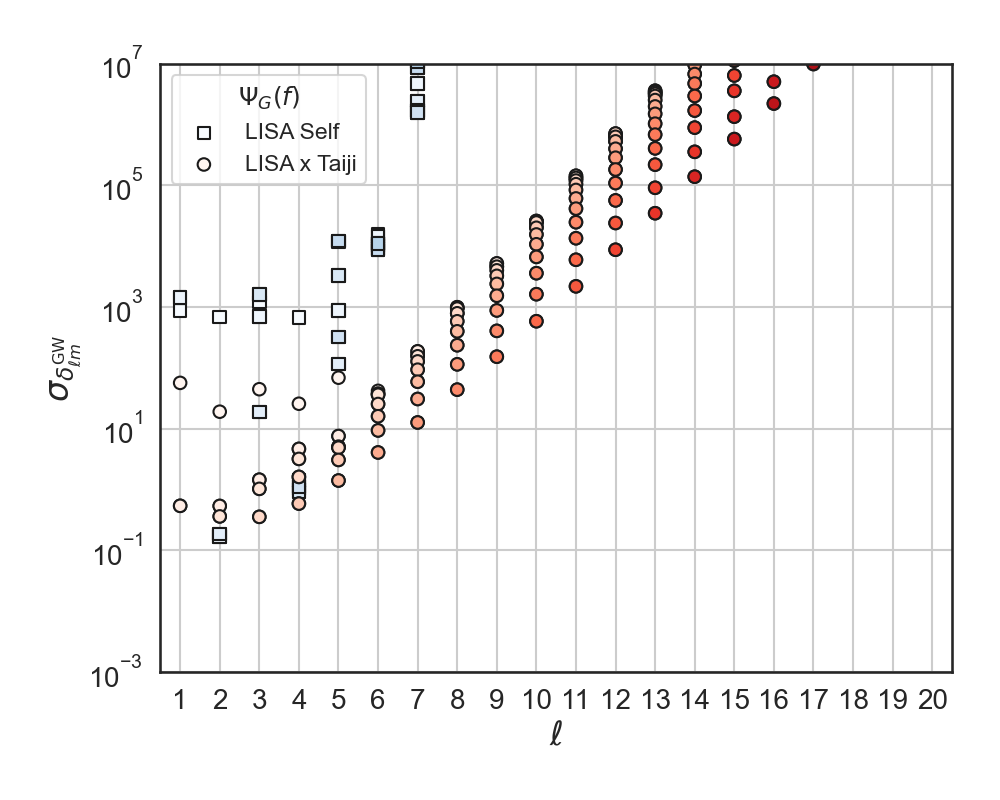}
\end{subfigure}
\caption{The expected (marginalized) forecast error of each relative anisotropic coefficient, $\delta_{\ell m }^{\rm GW}$, as observed by a network made of the three LISA satellites and the LISA-Taiji constellations. The first panel assumes the frequency dependence of the signal (cf.  (\ref{PowerLawOmega}))  $\psi(f)=\left(\frac{f}{f_0}\right)^{2/3}$, while the second is obtained with $\psi(f)=\psi_G(f)$ as in  \eqref{psi_galaxy}. A total observation time of $T=10$ yr and a fiducial value of $\hat\Omega_0=2\times 10^{-11}$ at the pivot frequency $f_0 = 1$~mHz are assumed.}
\label{fig:sigma_lm}
\end{figure}

	Following the same procedure adopted in \cite{Mentasti:2023gmg}, we perform a Fisher forecast on the measurement of the GW amplitude and anisotropies, by expanding the $\chi^2$ to quadratic order in the departure between their measured and their fiducial values. We obtain 
	\begin{align}\label{chi-squared-Fisher-map}
		\chi^2_{F} \equiv& (\Omega_0-\hat\Omega_0)^2\mathcal{F}_{\Omega_0\Omega_0} +2(\Omega_0-\hat\Omega_0)\sum_{\ell m}\mathcal{F}_{\Omega_0\delta_{\ell m}}(\delta^{GW}_{\ell m}-\hat\delta^{GW}_{\ell m}) \nonumber\\
		& +\sum_{\ell m,\ell'm'}(\delta^{GW}_{\ell m}-\hat\delta^{GW}_{\ell m}) \mathcal{F}_{\delta_{\ell m}\delta_{\ell'm'}} (\delta^{GW}_{\ell m}-\hat\delta^{GW}_{\ell' m'}) \;, 
	\end{align}
	where the Fisher matrix elements are 
	\begin{align}
		\mathcal{F}_{\alpha\beta} &\equiv \left. \frac 1 2\frac{\partial^2\chi^2_{\rm LT}}{\partial\alpha\partial\beta}\right|_{\substack{\alpha=\hat\alpha\\\beta=\hat \beta}} \;, 
	\end{align}
	and where $\hat \Omega_0$ and $\hat \delta^{GW}_{\ell m}$ are the fiducial values of the corresponding non-hatted quantities. The Fisher matrix elements turn out to be
	\begin{align}\label{eq:final_fisher}
		\mathcal{F}_{\Omega_0\Omega_0}= n \, \, I_{00,0} \;\;,\;\;\;\;\; 
		\mathcal{F}_{\delta_{\ell m}^{GW}\omega}=\mathcal{F}_{\Omega_0\delta_{\ell m}^{GW}}= n \, \frac{\hat\Omega_0}{\sqrt{4\pi}}I_{0\ell,0m'} \;\;,\;\;\;\; 
		\mathcal{F}_{\delta_{\ell m}^{GW}\delta_{\ell' m'}^{GW}}=n \, \frac{\hat\Omega_0^2}{4\pi}I_{\ell\ell', mm'} \;, 
	\end{align}
	where we defined
	\begin{align}\label{Ilm_COMPLETE_allmultipoles}
		I_{\ell \ell',mm'}=\frac 1 2\sum_{OO'}\left(\frac{3\left(\frac{H_0}{\text{Hz}}\right)^2}{4\pi^2\sqrt{4\pi}}\right)^2 \frac{T_s}{\text{s}}\int_{0}^{\infty}\frac{df}{\text{Hz}}\psi_G^2(f)\left(\frac{\text{Hz}}{f}\right)^6\frac{1}{N_t}\sum_{i=1}^{N_t}\frac{ \tilde R_{OO'}^{\ell m}(f,t_i)\tilde R_{OO'}^{\ell' m'*}(f,t_i)+\text{cc.}}{D_{OO'}(f)\,\text{Hz}^2} \;, 
	\end{align}
	and where for the numerical evaluations we chose $H_0=70\,\text{km}\,\text{s}^{-1}\text{Mpc}^{-1}=2.27 \times 10^{-18}\text{Hz}$. From the chi-squared in \eqref{chi-squared-Fisher-map} we can then define the posterior distribution of the energy density $\Omega_{0}$ and the anisotropic coefficients $\{\delta_{\ell m}\}$ as
	\begin{align}\label{posterior_map}
	P(\{\delta_{\ell m}\})=\frac{1}{\mathcal{N}}\pi\left(\Omega_{0},\{\delta_{\ell m}\}\right)e^{-\frac 1 2 \chi^2_F(\Omega_{0},\{\delta_{\ell m}\})}\,,
	\end{align}
where $\pi\left(\Omega_{0},\{\delta_{\ell m}\}	\right)$ is the prior distribution on our parameters, which we assume to be flat, and $\mathcal{N}$ the probability normalization factor.
In figure \ref{fig:sigma_lm} we show the forecast errors in the measurement of each coefficient $\delta^{GW}_{\ell m}$ after marginalizing the posterior distribution \eqref{posterior_map} over all the other parameters, in a similar way to what has been done in \cite{Mentasti:2023gmg}. Not surprisingly, a network made of LISA and Taiji would give better constraints in the measurement of the high $\ell$ anisotropic coefficients of the SGWB. Because of the longer baseline, the overlap functions of the instruments for each anisotropy up to $\ell=20$ have significant support at the frequencies where the LISA and Taiji will operate. On the other hand, for the lowest even multipoles ($\ell=2,4$) LISA-alone shows a better response.~\footnote{This is not contradictory, since, as mentioned above, the LISA-Taiji analysis includes only cross-correlations $\left\langle {\cal C}^{LT} \right\rangle$ across the two different constellations, see eq. \eqref{CLT}.} This is due to the fact that in the opposite limit, i.e. where the baseline between the instruments is short, the response to these specific multipoles is not suppressed~\cite{Mentasti:2020yyd}.

\section{Disentangling the galactic and the extragalactic signals}\label{sec:galaxy+back}
In the previous section, we studied the detectability of the anisotropies of the SGWB without any prior knowledge of them. More precisely, the galactic morphology outlined in appendix~\ref{app:galactic_deltalm} has been taken as a fiducial model, but its knowledge has not been assumed in the data analysis. In this section, we instead assume the specific anisotropic galaxy pattern determined by the galaxy, with the specific spatial distribution described in appendix~\ref{app:galactic_deltalm}, both in the fiducial model and as a prior knowledge in the analysis.  This means that, in the present analysis,  the coefficients $\delta_{\ell m}^{GW}$ in  \eqref{chi_sq_cross_general} have known values, and we assume that only the overall scale $\Omega_G$ of the galactic signal is unknown. We also assume the presence of an isotropic extragalactic component of unknown amplitude $\Omega_B$, so that the full signal is 
\begin{align}\label{OmegaGW_gal+extra}
		\Omega_{GW}(f,\hat n)=\Omega_G\,\psi_G(f)\sum_{\ell m}\delta^{GW}_{\ell m}Y_{\ell m}(\hat n) + \Omega_B\,\left(\frac{f}{f_0}\right)^{2/3},
	\end{align}
where $\psi_G(f)$ is the same functional frequency dependence adopted in the previous section, and, as already mentioned the coefficients $\delta_{\ell m}^{GW}$ are obtained from the spatial distribution given in appendix~\ref{app:galactic_deltalm}.
Finally, the frequency dependence of the extragalactic signal is the expected one, as also considered in \cite{Babak_2023}. With the assumption of perfect knowledge of the galaxy morphology, and of the frequency-dependence of the galactic and extragalactic components, the two only unknown quantities in the signal ({\ref{OmegaGW_gal+extra}) are the two amplitudes $\Omega_B$ and $\Omega_G$, which provide the fractional energy density contributions of the two components at the pivot scale $f_0=1$~mHz
~\footnote{A more general analysis can be performed assuming some specific functional form for the frequency dependencies. This will generally lead to more parameters and to a broader posterior distribution for the two amplitudes. This more general case, which is left for future work, is not conceptually different from the one presented here.}.
The chi-squared introduced in  \eqref{chi_sq_cross_general}, then becomes a function of these two parameters:
	\begin{align}
		\chi^2_{LT}&=\left(\frac{3H_0^2}{4\pi^2\sqrt{4\pi}}\right)^2 \sum_{OO'}\int_{0}^{\infty} df\frac{1}{f^6}\int_0^Tdt\frac{1}{D_{OO'}(f,t)}\times\nonumber\\
		&\times\left|\left(\Omega_B-\hat \Omega_B\right)\left(\frac{f}{f_0}\right)^{2/3}\sqrt{4\pi}\tilde R^{00}_{OO'}(f,t)+\left(\Omega_G-\hat\Omega_G\right)\psi_G(f)\sum_{\ell m}\delta^G_{\ell m}\tilde R^{\ell m}_{OO'}(f,t)\right|^2\,,\nonumber\\
\end{align}
where the denominator $D_{OO'}$ is formally defined as in ~(\ref{DOOp}),  in terms of the new function 
\begin{align}
\mathcal{T}_{OO'}(f,t)&\equiv\frac{3H_0^2}{4\pi^2\sqrt{4\pi}f^3}\left[\hat \Omega_B\sqrt{4\pi}\tilde R^{00}_{OO'}(f,t)+\hat \Omega_G\psi_G(f)\sum_{\ell m}\hat\delta_{\ell m}^{G}\tilde R^{\ell m}_{OO'}(f,t)\right]+\delta_{OO'}\tilde N_O(f)\,. 
\end{align}
Accordingly to \cite{10.1093/mnras/stac3686} and \cite{Babak_2023}, we set the fiducial values of the unknown parameters to $\hat\Omega_{G}=2\times 10^{-11}$ and $\hat \Omega_{B}=3.78\times 10^{-13}$, both defined at the pivot frequency $f_0=1$~mHz. We can further simplify the notation to 
\begin{align}
\label{Chisq_gal_extra_final}
		\chi^2_{LT}&=\sum_{OO'}\int_{0}^{\infty} df\int_0^Tdt\left|\left(\Omega_B-\hat \Omega_B\right)\tilde R^{B}_{OO'}(f,t)+\left(\Omega_G-\hat\Omega_G\right)\tilde R^{G}_{OO'}(f,t)\right|^2 \;, 
 \end{align}
by  introducing the dimensionless response functions (divided by the total variance)
\begin{align}\label{RB_timefrequency}
		\tilde R^{B}_{OO'}(f,t)&=\left(\frac{3H_0^2}{4\pi^2\sqrt{4\pi}f^3}\right)^2\frac{\sqrt{4\pi}}{D_{OO'}(f,t)}\left(\frac{f}{f_0}\right)^{2/3}\tilde R^{00}_{OO'}(f,t) \;, 
\end{align}
for the extragalactic, and 
\begin{align}\label{RG_timefrequency}
  		\tilde R^{G}_{OO'}(f,t)&=\left(\frac{3H_0^2}{4\pi^2\sqrt{4\pi}f^3}\right)^2\frac{\psi_G(f)}{D_{OO'}(f,t)}\sum_{\ell m}\delta_{\ell m}^GR^{\ell m}_{OO'}(f,t) \;, 
	\end{align}
for the galactic signal.
The Fisher approximation of the chi-squared (\ref{Chisq_gal_extra_final}) is 
	\begin{align}
		\chi^2_{F} \equiv (\Omega_B-\hat\Omega_B)^2\mathcal{F}_{\Omega_B\Omega_B} +2(\Omega_B-\hat\Omega_B)(\Omega_G-\hat\Omega_G)\mathcal{F}_{\Omega_B\Omega_G}+(\Omega_G-\hat\Omega_G)^2\mathcal{F}_{\Omega_G\Omega_G} \;, 
	\end{align}
	where the Fisher matrix elements are 
	\begin{align}\label{Fish_timefrequency}
		\mathcal{F}_{\Omega_B\Omega_B}&=\sum_{OO'}\frac{T_s}{\text{s}}\int_{0}^{\infty}\frac{df}{\text{Hz}}\frac{1}{N_t}\sum_{i=1}^{N_t}\left|\tilde R_{OO'}^{B}(f,t_i)\right|^2 \;,\nonumber\\
		\mathcal{F}_{\Omega_B\Omega_G}&=\frac 1 2\sum_{OO'}\frac{T_s}{\text{s}}\int_{0}^{\infty}\frac{df}{\text{Hz}}\frac{1}{N_t}\sum_{i=1}^{N_t}\left(\tilde R_{OO'}^{B}(f,t_i)\tilde R_{OO'}^{G*}(f,t_i)+\text{cc.}\right)\;,\nonumber\\
		\mathcal{F}_{\Omega_G\Omega_G}&=\sum_{OO'}\frac{T_s}{\text{s}}\int_{0}^{\infty}\frac{df}{\text{Hz}}\frac{1}{N_t}\sum_{i=1}^{N_t}\left|\tilde R_{OO'}^{G}(f,t_i)\right|^2\;.
	\end{align}

 \subsection{Frequency-only vs. time-frequency analysis}

The filter functions $\tilde Q$ adopted in the data analysis are weights that combine measurements obtained at different frequencies to maximize the signal-to-noise ratio (SNR) of the combination. Under the hypothesis of stationary signal and noise, and for the case of an isotropic signal, the optimal weights $\tilde Q$ are time-independent, as the same linear combination of the measured frequencies maximizes the SNR at all times. On the contrary, the weights $\tilde Q^{LL}_{OO'}(f,t)$ and $\tilde Q^{LT}_{OO'}(f,t)$ employed in our analysis, see eqs.~\eqref{CLL} and~\eqref{CLT}, depend also on time, accounting for the fact that the time positions of the satellites change with time, so that different linear combinations of the measured frequencies maximizes at different times the SNR associated to an anisotropic signal (that is measured differently at different times even for a stationary signal, due to the time-dependence of the satellite response functions).  

A simpler analysis can be performed by ignoring this time information and performing a frequency-only analysis, in which essentially data obtained at any given frequency are averaged over the year. This will result in a sub-optimal estimator. We want to assess the improvement of the optimal time-frequency analysis done above vs. this simplified frequency-only analysis.

The frequency-only analysis is performed by integrating along the whole time of observation the cross-correlators between the different channels (and instruments in the case of the LISA-Taiji network) first, and by then applying the filters $\tilde Q_{OO'}^{LL}(f)$ and $\tilde Q_{OO'}^{LT}(f)$, which are now functions of the frequency only. Concretely, this means that eqs.~\eqref{CLL} and~\eqref{CLT} are replaced by
	\begin{align}\label{CLLT_wrong}
		\mathcal{C}^{LL} &\equiv \sum_{O, O^{\prime}} \int_{-\infty}^{+\infty} d f \,\tilde{Q}_{O O^{\prime}}(f)\int_0^T d t\left[\tilde{m}^L_O(f, t) \tilde{m}^{L,*}_{O^{\prime}}(f, t)-\left\langle\tilde{n}^L_O(f, t) \tilde{n}^{L,*}_{O^{\prime}}(f, t)\right\rangle\right] \;,\nonumber\\
		\mathcal{C}^{LT} &\equiv \sum_{O, O^{\prime}} \int_{-\infty}^{+\infty} d f\,\tilde{Q}^{LT}_{O O^{\prime}}(f)\int_0^T d t \, \tilde{m}^L_O(f, t) \tilde{m}^{T,*}_{O^{\prime}}(f, t)\,.
	\end{align}
We can now proceed along the lines of the analysis performed in section \ref{sec:galaxy+back} by introducing the time-averaged detector overlap functions
\begin{align}
		\tilde R^{\ell m}_{OO'}(f)&\equiv\frac{1}{T}\int_0^T dt \tilde R^{\ell m}_{OO'}(f,t) \;, 
\end{align}
and by employing them in the expressions 
	\begin{align}
		\mathcal{T}_{OO'}(f)&=\frac{3H_0^2}{4\pi^2\sqrt{4\pi}f^3}\left[\hat \Omega_B\sqrt{4\pi}\tilde R^{00}_{OO'}(f)+\hat \Omega_G\psi_G(f)\sum_{\ell m}\hat\delta_{\ell m}^{G}\tilde R^{\ell m}_{OO'}(f)\right]+\delta_{OO'}\tilde N_O(f) \,.
	\end{align}
The optimal chi-squared
of this frequency-only analysis (which is sub-optimal when compared to the one computed in the previous section) acquires the form 
	\begin{align}\label{Chisq_gal_extra_final_wrong}
		\chi^2_{LT}&=\sum_{OO'}\int_{0}^{\infty} df\left|\left(\Omega_B-\hat \Omega_B\right)\tilde R^{B}_{OO'}(f)+\left(\Omega_G-\hat\Omega_G\right)\tilde R^{G}_{OO'}(f)\right|^2 \;, 
\end{align}
where the coefficients associated with the extragalactic and the galactic term are, respectively, 
\begin{align}\label{RG_frequencyonly}
		\tilde R^{B}_{OO'}(f)&=\left(\frac{3H_0^2}{4\pi^2\sqrt{4\pi}f^3}\right)^2\frac{\sqrt{4\pi}}{D_{OO'}(f)}\left(\frac{f}{f_0}\right)^{2/3}\tilde R^{00}_{OO'}(f) \;, \nonumber\\
		\tilde R^{G}_{OO'}(f)&=\left(\frac{3H_0^2}{4\pi^2\sqrt{4\pi}f^3}\right)^2\frac{\psi_G(f)}{D_{OO'}(f)} \sum_{\ell m}\delta_{\ell m}^G R^{\ell m}_{OO'}(f) \;, 
	\end{align}
and where the denominators $D_{OO'}(f)$ are related to the functions $\mathcal{T}_{O'O}(f)$ analogously to ~(\ref{DOOp}), and the Fisher matrix elements are in this case 
	\begin{align}\label{Fish_frequencyonly}
		\mathcal{F}_{\Omega_B\Omega_B}&=\sum_{OO'}\frac{T_s}{\text{s}}\int_{0}^{\infty}\frac{df}{\text{Hz}}\left|\tilde R_{OO'}^{B}(f)\right|^2 \;,\nonumber\\
		\mathcal{F}_{\Omega_B\Omega_G}&=\frac 1 2\sum_{OO'}\frac{T_s}{\text{s}}\int_{0}^{\infty}\frac{df}{\text{Hz}}\left(\tilde R_{OO'}^{B}(f)\tilde R_{OO'}^{G*}(f)+\text{cc.}\right)\;,\nonumber\\
		\mathcal{F}_{\Omega_G\Omega_G}&=\sum_{OO'}\frac{T_s}{\text{s}}\int_{0}^{\infty}\frac{df}{\text{Hz}}\left|\tilde R_{OO'}^{G}(f)\right|^2\;.
	\end{align}
We stress that the chi-squared (\ref{Chisq_gal_extra_final_wrong}) differs from \eqref{Chisq_gal_extra_final} from the fact that here we take an average over the time of the response functions first, while in the previous analysis, we built a full measurement for each time window we considered. As we already remarked, and as we show with one explicit example in figure \ref{fig:fisher_contours}, this simplified method presented here results in a less accurate determination of the model parameters. 

 \subsection{The posterior distribution and final forecast}
 
 In a similar way to what we did in the previous section, we introduce the posterior probability of the two parameters of our problem, $\Omega_G$ and $\Omega_B$:
\begin{align}\label{posterior_galaxy}
P(\Omega_{G},\Omega_B)=\frac{1}{\mathcal{N}}\pi(\Omega_{G},\Omega_B)e^{-\frac 1 2 \chi^2_F(\Omega_{G},\Omega_B)}\,,
\end{align}
where also in this case a flat prior $\pi(\Omega_{G},\Omega_B)$ is assumed and where 
the chi-squared is defined in  \eqref{Chisq_gal_extra_final} or \eqref{Chisq_gal_extra_final_wrong} depending if we want to adopt a time-frequency or a frequency-only analysis. In figure \ref{fig:fisher_contours} we show the contour levels of \eqref{posterior_galaxy} for a total observation time of $T=10$ years with either the LISA constellation alone or the LISA-Taiji networks (in this second case, we do not include correlations between two measurements performed within satellites of the same constellation). We note from the figure that, at face value, LISA alone would perform better than when combined with Taiji. However, we stress that any SGWB search performed by LISA alone provides a perfectly unbiased estimator of the signal (in  \eqref{CLL}) only if a perfect knowledge of the noise statistics is assumed. On the other hand, the cross-correlators between LISA and Taiji TDIs as in  \eqref{CLT} are not noise-correlated. 

We furthermore note that a time-frequency analysis would not make significant improvements with respect to a frequency-only analysis for LISA alone (for LISA-self, the contours of the frequency-only and of the time-frequency analysis in Figure \ref{fig:fisher_contours} are practically superimposed to each other, and cannot be distinguished by eye), while there is a significant improvement for a LISA-Taiji joint analysis. The reason is the fact that we assumed an isotropic extragalactic SGWB in \eqref{OmegaGW_gal+extra}, for which the response functions of LISA do not change with time. The same is not true for the LISA-Taiji analysis, due to the variation of the relative orientation of LISA and Taiji arms with respect to their baseline. This also increases the time variation of the LISA-Taiji response function to the galactic signal, with respect to the LISA-self case. In appendix~\ref{app:ftvsfLISA} we verify that indeed, for the LISA-self case, the time-dependence of the galactic signal is not strong enough to make the time-frequency analysis significantly better than the sub-optimal frequency-only one. 

\begin{figure}[h!]
		\centerline{
			\includegraphics[width=0.9\textwidth,angle=0]{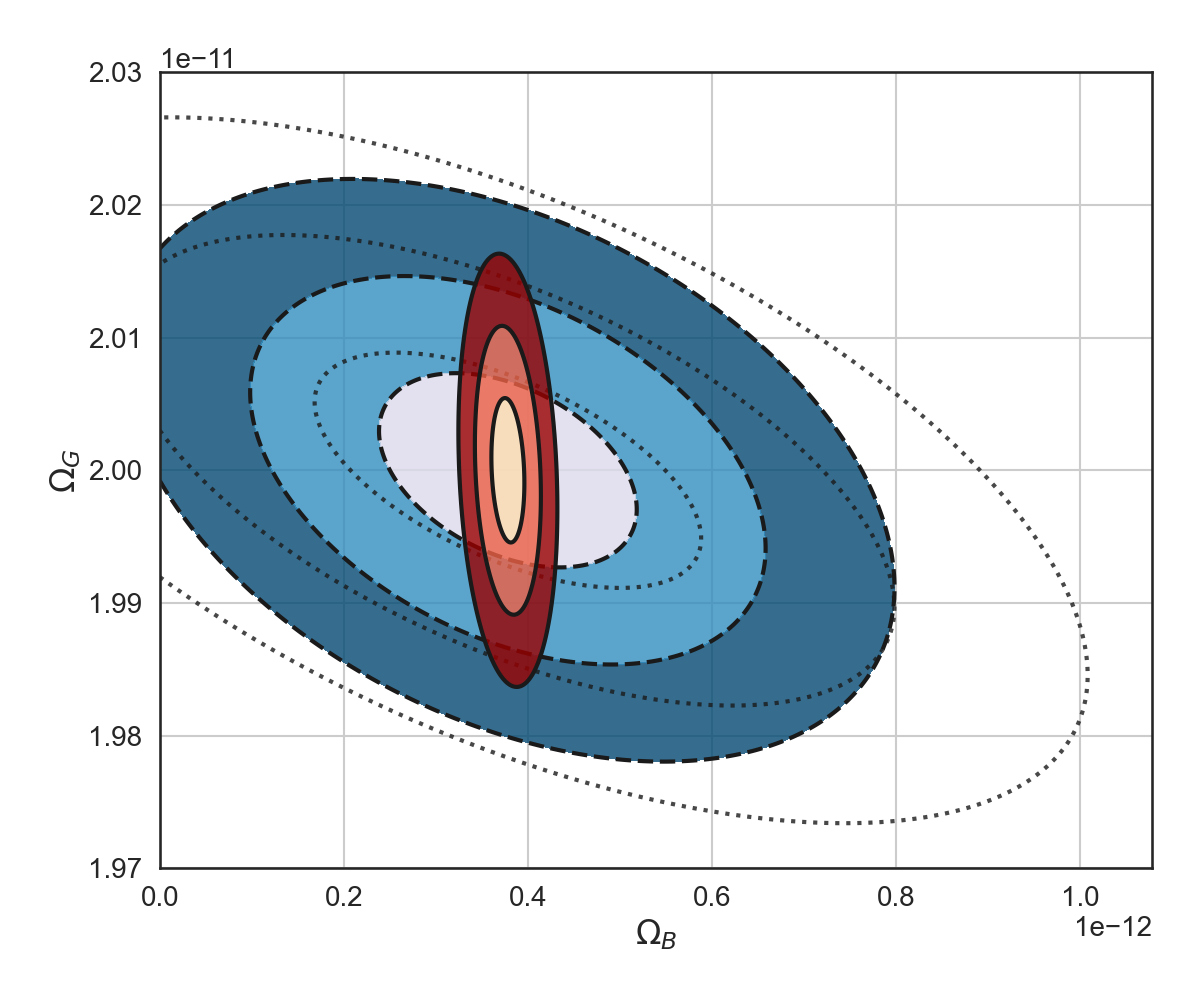}
		}		\caption{Contour lines at $1\sigma$, $2\sigma$ and $3\sigma$ levels of the posterior distribution of  \eqref{posterior_galaxy}. The red regions, with thick lines, correspond to the forecast posterior distribution of a measurement performed by the LISA constellation alone. The blue regions, with dashed lines, represent the forecast obtained using a network made of LISA and Taiji, without including self-correlations within each constellation. This second analysis provides a less precise determination of the model parameters, but it is not biased by imperfect knowledge of the instrumental noise.  For comparison, a frequency-only analysis for the LISA-Taiji network leads to the contour levels displayed with dotted lines. For LISA-self, the frequency-only analysis produces contours superimposed to the solid ones. A time of observation of $T=10$ yr and fiducial value of $\hat\Omega_G=2\times 10^{-11}$, $\hat\Omega_B=3.78\times 10^{-13}$ are assumed, as expected from \cite{10.1093/mnras/stac3686} and \cite{Babak_2023} (as discussed in the text).}
		\label{fig:fisher_contours}
	\end{figure}
	
\section{Conclusions}\label{sec:conclusions}

In the present work, we applied the formalism developed in \cite{Mentasti:2023gmg} and \cite{Bartolo_2022} to produce forecasts on the measurement of a stochastic GW background with two possible configurations of space-based detectors.
We considered two possible analyses, under the assumption of a stationary SGWB in its reference frame, for both a network made of the three LISA satellites and a second network where LISA and Taiji can operate jointly.
In the first analysis, we produced a fiducial model of the distribution of the GW galactic signal and we took advantage of their anisotropic distribution in the reference frame of the Earth to forecast at which level of accuracy the sky maps of these sources can be obtained.
We realized that, because of the fact that the GW emission by the unresolved galactic binaries interests just the lowest frequencies detectable by LISA and Taiji (as in  \eqref{psi_galaxy}), only the very first multipoles $\{\delta^{GW}_{\ell m}\}$ of the anisotropic distribution of the sources are resolvable.
This is due to the fact that the higher multipoles overlap functions of the instruments, as in  \eqref{R_oop_lm} have their maximum at higher values of frequencies, where there are no unresolved galactic sources emitting.
In this case, the network of LISA and Taiji outperforms LISA alone thanks to the fact that the baselines between the LISA and the Taiji satellites are greater than the distance between each pair of LISA satellites.
This is clear from figure \ref{fig:sigma_lm}, where the forecast error $\sigma_{\delta^{GW}_{\ell m}}$ represents the accuracy in determining the value of the anisotropic coefficient $\delta^{GW}_{\ell m}$ as introduced in \eqref{eqPOmega}.
In general, the network made of LISA and Taiji displays a better sensitivity to the anisotropic coefficients up to $\ell=20$ with respect to LISA alone, because of the longer baseline between the instruments. On the other hand, LISA alone has an enhanced response to the monopole and the first even $\ell$ value multipoles, since the corresponding overlap functions are not suppressed in the low-baseline limit.
Since the coefficients $\delta^{GW}_{\ell m}$ are of the order of $O(10^{-1})$ or smaller, it means that a high-resolution galaxy map of the galactic background with our space-based detectors is difficult to obtain.
These forecast errors decrease with the square root of the total time of observation and with the squared of the power spectral density of the noise (since the data stream is almost noise-dominated).

From the results displayed in figure \ref{fig:sigma_lm} we see that it is crucial to adopt the real model of the galaxy power spectrum as a function of the frequency, i.e. after the removal of the resolvable sources. This removal subtracts power from the SGWB at comparatively higher frequencies, thus worsening the detectability of the anisotropic dependence of the residual SGWB. While the lower panel of figure \ref{fig:sigma_lm} is a better representation of the detectability of the SGWB of galactic origin, the results in the upper panel can be used as a proxy on the detectability of an extra-galactic component, for which individual high-frequency sources cannot be removed.~\footnote{This is not an exact representation of the study of an extra-galactic $\propto f^{2/3}$ component, since the denominator of the SNR is affected by the variance of the signal of the fiducial galactic model that we have adopted; however, it is a good approximation, since the denominator is dominated by the variance of the noise in most of its frequency range.}

Another novelty of this analysis is the fact that, for the first time, we have a precise forecast of the measurement of the anisotropic coefficients $\delta^{GW}_{\ell m}$ separately. For example, the study of the anisotropy presented in \cite{Bartolo_2022} does not provide a precise forecast for the measurement of the specific coefficients $\delta^{GW}_{\ell m}$, but it estimates
the level of their detectability under the assumption of statistical isotropy.  Under this assumption, ref. \cite{Bartolo_2022} provided an estimate of the level of detection of the combination $C_\ell^{GW} \propto \sum_m \left\vert \delta^{GW}_{\ell m} \right\vert^2$. As shown in \cite{Bartolo_2022}, the LISA response function for this combination is isotropic, and therefore the motion of the LISA satellites is completely irrelevant to that assessment. We, on the contrary, have studied the determination of the specific anisotropic coefficients $\delta^{GW}_{\ell m}$, for which the precise time-dependent orientation of the LISA constellation must be known. The same is true when Taiji is added to the analysis. For this reason, we adopted the model for the satellite trajectories presented in appendix~\ref{app:positions}, 
and we then computed the overlap functions of the LISA and Taiji networks as a function of time.

As a second analysis, we treated two possible sources of SGWB radiation together and we forecasted the ability of LISA and LISA+Taiji to disentangle their separate contributions.
We considered a galactic background as in the previous analysis. However, in this case, we assumed perfect prior knowledge of the frequency dependence of the power spectrum and its anisotropic distribution in the data analysis, keeping as the only unknown the overall logarithmic energy density, i.e. the parameter $\Omega_G$ in  \eqref{OmegaGW_gal+extra}.
On top of the galactic signal, we assumed the presence of an isotropic extragalactic component, of astrophysical nature, due to the superposition of gravitational waves sourced by a large number of compact binaries~\cite{Babak_2023}. We assumed the typical $f^{2/3}$ frequency dependence of this signal, keeping also in this case, a free normalization $\Omega_B$ in  \eqref{OmegaGW_gal+extra}. In figure \ref{fig:fisher_contours} we show the posterior distribution of the parameters $\Omega_G$ and $\Omega_G$, centered in the value predicted by our fiducial model.
The result appears to indicate that LISA alone performs better than when correlated with Taiji. This is due to the fact that the main contributions to the data stream of both the sources of GW background radiation come from the lowest multipoles, at which LISA alone is more sensitive than when correlated with Taiji.
However, we stress the fact that a measurement performed with LISA alone is perfectly unbiased only under the ideal condition of perfect knowledge of the noise of the instrument: any error in the estimate of the noise budget would translate into a bias in the final estimate.
On the other hand, the cross-correlation of LISA and Taiji produces an unbiased estimator of the GW power spectrum, as the noises of the two constellations are not cross-correlated.

A final important point we would like to stress is the advantage of performing a time-frequency analysis compared to the traditional frequency domain in the case of a network made of multiple instruments.
As it appears evident from the figure \ref{fig:fisher_contours}, a network made of LISA and Taiji will produce stronger constraints when time-frequency response functions are employed (the thick lines of the plot) with respect to the frequency-only analysis (the dashed lines of the plot). This is mainly due to the relative orientations of the baseline and the arms of the two instruments, which results in a very complex time dependence of the response (overlap) functions of the network.
On the other hand, when LISA alone is considered, the response (overlap) functions have a dominant constant plus a subdominant time-dependent component which is very regular over time (see appendix \ref{app:ftvsfLISA} for more details). Assuming that data are continuously taken over the year (or, more realistically, that gaps in the data are evenly distributed across the orbit), one finds that the analysis in the frequency-only domain (with time-averaged response functions) gives only marginally worse results than the optimal time-frequency analysis.

\acknowledgments 

G.M. acknowledges support from the Imperial College London Schr\"odinger Scholarship scheme. C.R.C. acknowledges support under a UKRI Consolidated Grant ST/T000791/1. M.P. acknowledges support by Istituto Nazionale di Fisica Nucleare (INFN) through the Theoretical Astroparticle Physics (TAsP) project and by the MIUR PRIN Bando 2022 - grant 20228RMX4A. We thank Mauro Pieroni for the useful discussions and Francesco Mentasti for the graphics of figure \ref{fig:LISA-Taiji-scheme}.
	
	
	\appendix
	
	\section{Position of the satellites}\label{app:positions}
	
	We assume a circular orbit of the Earth around the Sun, of radius $R$ equal to one astronomical unit ($R \simeq 1.496 \times 10^8 \, {\rm km}$). We consider a Cartesian coordinate system with the Sun at its center and with the $xy$ plane coinciding with the ecliptic plane. We align the $x$ axis with the position of the Earth at the reference time $t=0$. The position of the Earth in this coordinate system is therefore given by 
	\begin{equation}
		x_E^i  \left( t \right) = \left( {\cal R}_I \left[ \omega \, t \right] \right)^i_j \, y_E^j \, R \;,
	\end{equation} 
	where $\omega = \frac{2 \pi}{1 \, {\rm year}}$ is the angular frequency of the orbit, and where
	\begin{equation}
		{\cal R}_I \left[ \theta \right] = \left( 
		\begin{array}{ccc}
			\cos \left( \theta \right) & -\sin \left( \theta \right) & 0 \\ 
			\sin \left( \theta \right) & \cos \left( \theta \right) & 0 \\ 
			0 & 0 & 1 
		\end{array} 
		\right) \;\;\;,\;\;\; \vec{y}_E = \left( \begin{array}{c} 1 \\ 0 \\ 0 \end{array} \right) \;.
	\end{equation} 
	To describe the position of the satellites of a given constellation (LISA or Taiji) in this coordinate system, we introduce a second rotation matrix 
	\begin{equation}
		{\cal R}_{II} \left[ \theta \right] = \left( 
		\begin{array}{ccc}
			\frac{3 + \cos \theta}{4} & \frac{\sin \theta}{2}  & - \frac{\sqrt{3} \left( 1 - \cos \theta \right)}{4} \\ 
			-\frac{\sin \theta}{2} & \cos \theta  & - \frac{\sqrt{3}}{2} \sin \theta \\ 
			- \frac{\sqrt{3} \left( 1 - \cos \theta \right)}{4} &  \frac{\sqrt{3}}{2} \sin \theta & \frac{1+3 \cos \theta}{4} 
		\end{array} 
		\right) \;, 
	\end{equation} 
	and the tree unit vectors 
	\begin{equation}
		{\hat v}_1 = \left( \begin{array}{c} - \frac{1}{2} \\ 0 \\ - \frac{\sqrt{3}}{2} \end{array} \right) \;\;,\;\; 
		{\hat v}_2 = \left( \begin{array}{c}  \frac{1}{4} \\ - \frac{\sqrt{3}}{2}  \\  \frac{\sqrt{3}}{4} \end{array} \right) \;\;,\;\; 
		{\hat v}_3 = \left( \begin{array}{c}  \frac{1}{4} \\ \frac{\sqrt{3}}{2}  \\  \frac{\sqrt{3}}{4} \end{array} \right) \;\;.  
		\label{vectors}
	\end{equation} 
In terms of these quantities, the position of the three satellites is given by 
	\begin{eqnarray} 
		x_\lambda^i &=& \left( {\cal R}_I \left[ \omega \, t  + \alpha_c \right] \right)^i_j 
		\left[ R \, y_E^j + \frac{L}{\sqrt{3}} \left( {\cal R}_{II} \left[ \omega \, t + \beta_c \right] \right)^j_k {\hat v}_\lambda^k \right]  + {\rm O } \left( \frac{L^2}{R} \right) \;,
		\label{implicit-pos}
	\end{eqnarray} 
	where the Greek index $\lambda = 1,\, 2 ,\, 3$ refers to one of the three satellites in the constellation, while the Latin indices (which also run from $1$ to $3$) are coordinate indices. The two constant phases $\alpha_c$ and $\beta_c$ are specific to each constellation. The expression (\ref{implicit-pos}) can be written more explicitly in the form 
	\begin{eqnarray}
		\vec{x}_1 &=& \Bigg\{ R \, \cos \left( \omega t + \alpha_c \right) + \frac{L}{4 \sqrt{3}} \left( \cos \left( 2 \omega t + \alpha_c + \beta_c \right) - 3 \cos \left( \alpha_c - \beta_c \right) \right) , \nonumber\\ 
		&& \;\;\;\; R \, \sin \left( \omega t + \alpha_c \right) + \frac{L}{4 \sqrt{3}} \left( \sin \left( 2 \omega t + \alpha_c + \beta_c \right) - 3 \sin \left( \alpha_c - \beta_c \right) \right) , \nonumber\\
		&& \;\;\;\; - \frac{L}{2} \cos \left( \omega t + \beta_c \right) \Bigg\} + {\rm O } \left( \frac{L^2}{R} \right) \;, 
		\nonumber\\ 
		\vec{x}_2 &=& {\rm same \; as } \; \vec{x}_1 \; {\rm but \; with \; } \beta_c \to \beta_c - \frac{2 \pi}{3} \;, \nonumber\\ 
		\vec{x}_3 &=& {\rm same \; as } \; \vec{x}_1 \; {\rm but \; with \; } \beta_c \to \beta_c + \frac{2 \pi}{3} \;. 
		\label{explicit-pos}
	\end{eqnarray}
The implicit relation  (\ref{implicit-pos}) manifestly shows that the motion of each constellation can be thought of as the composition of two rotations, both of periods equal to one year. The rotation $R_I \left[ \omega \, t  + \alpha_c \right]$ keeps the constellation at fix relative phase $\alpha_c$ with respect to the Earth.~\footnote{LISA is planned to orbit $20^\circ$ 
		behind the Earth, namely $\alpha_{\rm LISA} = - \frac{\pi}{9}$. The proposed orbit for 
		Taiji is of $20^\circ$ ahead of the Earth, namely $\alpha_{\rm Taiji} = + \frac{\pi}{9}$. The value of $\beta_c$ depends on the choice of the reference time, and we are not aware of any choice of the relative values of $\beta_c$ of the two constellations. In this work, we assume $\beta_{\rm LISA}=0$ and $\beta_{\rm Taiji}=1$ for definiteness.} Simultaneously, the three satellites rotate with 
	$R_{II} \left[ \omega \, t + \beta_c \right]$, which is a rigid rotation  (known as  cartwheel rotation) about the normal ${\hat n}_c$, of components 
	\begin{equation}
		n_c^i \equiv \left( R_I \left[ \omega \, t  + \alpha_c \right] \right)^i_j 
		\left( \begin{array}{c} - \frac{\sqrt{3}}{2} \\ 0 \\ \frac{1}{2} \end{array} \right)^j \;.
		\label{normal} 
	\end{equation} 
We see from this relation that the normal forms a constant $60^\circ$ angle with the $z-$axis, namely the plane formed by the three satellites is inclined of 60$^\circ$ with respect to the ecliptic plane (this is true for both LISA and Taiji). 
Ideally, one wishes to keep the three satellites of each constellation at the vertices of an equilateral triangle. While this is not exactly possible for elliptical orbits about the Sun, an approximate equilateral triangular configuration can be maintained at all times provided that the side of this triangle is much smaller than one astronomical unit. From ~(\ref{explicit-pos}), the three satellites are at the distances 
	\begin{equation}
		\left\vert \vec{x}_\lambda - \vec{x}_{\lambda'} \right\vert = L + {\rm O } \left( \frac{L^2}{R} \right) \;\;,\;\; \lambda \neq \lambda' \;, 
	\end{equation} 
	which is indeed constant to leading order. To prove that the orbits given above are indeed elliptical, let us compute the distances between the three satellites and the Sun (placed at the origin of our coordinate system) following from (\ref{explicit-pos}),  %
	\begin{eqnarray} 
		\left\vert \vec{x}_1 \right\vert &=& R - \frac{L}{2\sqrt{3}} \cos \left( \omega t + \beta_c \right) + {\rm O } \left( \frac{L^2}{R} \right) \;, \nonumber\\ 
		\left\vert \vec{x}_2 \right\vert & = & R - \frac{L}{2\sqrt{3}} \cos \left( \omega t + \beta_c - \frac{2 \pi}{3} \right) + {\rm O } \left( \frac{L^2}{R} \right) \;, \nonumber\\ 
		\left\vert \vec{x}_3 \right\vert & = & R - \frac{L}{2\sqrt{3}} \cos \left( \omega t + \beta_c + \frac{2 \pi}{3} \right) + {\rm O } \left( \frac{L^2}{R} \right) \;, 
	\end{eqnarray} 
	and let us compare them with the parametric relation for an ellipse in radial $\left\{ r ,\, \phi \right\}$ coordinates having a focus in the origin, 
	\begin{equation}
		r = \frac{R}{1+e \cos \left( \phi + \phi_0 \right)} = R \left[ 1 - e \, \cos \left( \phi + \phi_0 \right) + {\rm O } \left( e^2 \right) \right] \;,  
	\end{equation} 
	where the second expression assumes a small eccentricity $e \ll 1$, and where $\phi_0$ is a fixed angle. The comparison shows that the nearly fixed distance $L$ between the satellites is obtained by placing them in elliptical orbits with a small eccentricity of
	\begin{equation} 
		e = \frac{L}{2 \sqrt{3} \, R} \ll 1 \;. 
	\end{equation} 
	The designed arm length for LISA is $L^L = 2.5 \times 10^6 \, {\rm km}$, while the proposed arm length for Taiji is $L^T = 3 \times 10^6 \, {\rm km}$, corresponding to the two small eccentricities 
	\begin{equation}
		e_{\rm LISA} \simeq 0.00486 \;\;,\;\; e_{\rm Taiji} \simeq 0.00583 \;. 
	\end{equation} 
	
	\section{Model of the galactic SGWB}\label{app:galactic_deltalm}
	Using the results above, we can examine how well space-based networks can reconstruct the anisotropic gravitational-wave sky. The strongest stochastic and anisotropic signal at mHz frequencies is expected to be that of unresolved galactic binaries \cite{10.1093/mnras/stac3686,PhysRevD.104.043019}.
	We build a template morphology for the gravitational wave signal of galactic binaries using a bulge and disc stellar model of the galaxy \cite{Nelemans:2003ha,PhysRevD.86.124032}. The stellar density, as a function of galactocentric cartesian coordinates, is given by 
	\begin{align}
		\rho(x,y,z) = \rho_0\left[Ae^{-r^2/R^2_b}+(1-A)e^{-u/R_d}\sech^2(z/Z_d) \right]\,,
	\end{align}
	where $r^2=x^2+y^2+z^2 \equiv u^2+z^2$, $R_b$ is the characteristic radius of the bulge, and $R_d$ and $Z_d$ are the characteristic radius and scale height of the disc respectively. $A$ gives the relative weight of the stellar density in the bulge compared to the disc. The normalization of the stellar density $\rho_0$ is set to unity as we are only interested in the morphology of the signal. We set $A=0.25$, $R_b=500$ pc, $R_d=2500$ pc, and $Z_d=200$ pc.
	
	We form a template of the gravitational-wave sky by integrating the density in logarithmic intervals along all lines-of-sight $l_{\mathbf{\hat p}}$ corresponding to each {\tt Healpix} pixel $p$ on a map of resolution $N_{\rm side}=32$ with pixels of angular size 1.8$^\circ$ and angular Nyquist scale $\ell\approx 120$.  
 
	\begin{align}\label{map_Galaxy}
		{\cal M}(\mathbf{\hat p}) = \int_{l_{\mathbf{\hat p}}}\, \rho(\mathbf r) \,\frac{dl}{l}\,,
	\end{align}
	where the origin of each line-of-sight is set to the position of the sun in the galactocentric coordinate system, $[-8.1,0,0]$ kpc, and it is extended to 20 kpc in all directions around the sun location. The resulting map is shown in Figure~\ref{fig:galaxy}.
	
	Given the Fisher matrix estimates of \eqref{eq:final_fisher} for the relative anisotropy $\delta_{\ell m}^{\rm GW}$ and the spectral coefficients of the unit normalized template $a_{\ell m}^{\cal M}$, we can produce realizations of the `observed' sky that account for noise levels and mode coupling induced by each network of baselines. To generate the realization of the noise with the correct correlation structure, we generate a set of random, uncorrelated normal variates $\xi_{\ell m}$ with unit variance and rotate to a set $a_{lm}^N $ using the hermitian square root of the inverse Fisher matrix
	\begin{align}
		a_{lm}^N = {\cal F}^{-1/2}_{\ell m, \ell' m'} \xi^{\phantom{0}}_{\ell' m'}\,.
	\end{align}
	
	\begin{figure}[h!]
		\centerline{
			\includegraphics[width=0.9\textwidth,angle=0]{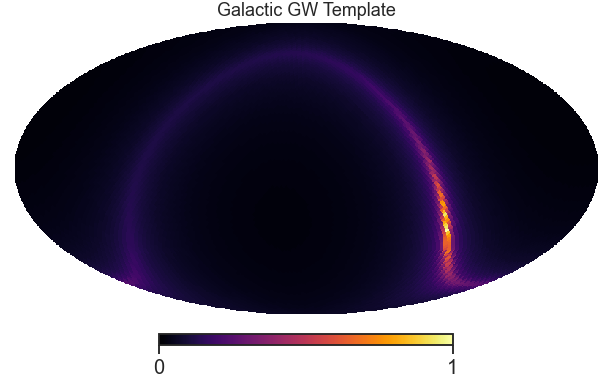}
		}
		\caption{Mollweide projection of the template morphology for the galactic GW signal arising from unresolved binaries. The template is unit normalized and in celestial coordinates.}   
		\label{fig:galaxy}
	\end{figure}

\section{Time-frequency vs. frequency only analysis for LISA-self} \label{app:ftvsfLISA}
Figure \ref{fig:fisher_contours} indicates that, for the analysis performed by the LISA constellation alone, the time-frequency domain does not bring a significant advantage with respect to the frequency-only analysis. This Appendix discusses the reason for this.
Firstly, let us consider the signal from the galaxy, which, due to its angular distribution  displayed in figure \ref{fig:galaxy}, induces a time-frequency response function \eqref{RG_timefrequency} that is well fit by  
\begin{align}
\tilde R^{G}_{OO'}(f,t)\simeq G_{OO'}(f)\left(1+A\cos(\frac{4\pi t}{T_e})\right) \;, 
\label{fit}
\end{align}
where $G_{OO'}(f)$ is a function of the frequency only and $A$ is the relative amplitude of the modulation of the galactic signal in the orbit of LISA as a function of time.
\begin{figure}[h!]
\centerline{
\includegraphics[width=0.7\textwidth,angle=0]{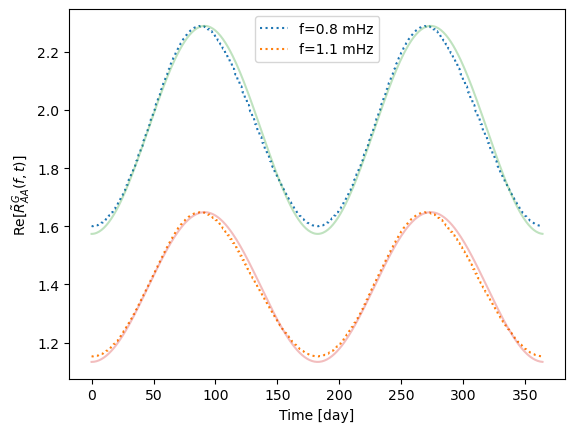}
}
\caption{Time evolution of the response function 
$\tilde R_{AA}^G(f,t)$ during one year, for two specific frequencies chosen as an example. Only the real part is shown, as the imaginary part is sub-dominant. The points are the numerical values used in our analysis (taken with a time interval of one day), while the solid line is the fitting function \eqref{fit}, with $A = 0.185$ in both cases.}
\label{fig:t-modulation-G}
\end{figure}
Figure~\ref{fig:t-modulation-G} shows the goodness of the fit for two frequencies in the $A$ channel. We note that the modulation is the same for all frequencies (namely, $A$ is constant, and no frequency-dependent phase is present in the argument of the cosine) as a consequence of the factorization of the angular and frequency dependencies in the galactic signal in eq.~\eqref{OmegaGW_gal+extra}. From our numerical fit, we obtain $A \simeq 0.185$, namely the amplitude of the modulated term accounts for about $20\%$ of the overall amplitude.
The frequency-only galactic response function, as introduced in the second line of eq. \eqref{RG_frequencyonly} for the sub-optimal analysis then evaluates to 
\begin{align}
\tilde R^{G}_{OO'}(f)&\equiv\frac{1}{T}\int_0^T dt \tilde R^{G}_{OO'}(f,t)\simeq G_{OO'}(f)\frac{1}{T_e}\int_0^{T_e}dt\left(1+A\cos(\frac{4\pi t}{T_e})\right)=\nonumber\\
&=G_{OO'}(f)\left(1+A\int_0^1dx\cos(4\pi x)\right)=G_{OO'}(f)\,, 
\end{align}
namely, the term that encodes the modulation cancels out in the average (this assumes measurements that are taken homogeneously across the yearly revolution). Secondly, we note that the response functions to the isotropic extragalactic background to be used in the time-frequency and in the sub-optimal frequency-only analysis (see eq.~\eqref{RB_timefrequency} and the first line of eq.~\eqref{RG_frequencyonly}) coincide, namely 
\begin{align}
\tilde R^{B}_{OO'}(f,t)=\tilde R^{B}_{OO'}(f)\,.
\end{align}
The final forecasts displayed in Figure \ref{fig:fisher_contours} depend on the Fisher matrices, introduced in eq.~\eqref{Fish_timefrequency} for the time-frequency analysis and in eq.~\eqref{Fish_frequencyonly} for sub-optimal frequency-only analysis, respectively. From the expressions written in this Appendix, the elements of the Fisher matrix for the time-frequency analysis acquire the form  
\begin{align}
\mathcal{F}_{\Omega_B\Omega_B}&=\sum_{OO'}\frac{T_s}{\text{s}}\int_{0}^{\infty}\frac{df}{\text{Hz}}\frac{1}{N_t}\sum_{i=1}^{N_t}\left|\tilde R_{OO'}^{B}(f,t_i)\right|^2= \sum_{OO'}\frac{T_s}{\text{s}}\int_{0}^{\infty}\frac{df}{\text{Hz}}\left|\tilde R_{OO'}^{B}(f)\right|^2\;,\nonumber\\
\mathcal{F}_{\Omega_B\Omega_G}&=\frac 1 2\sum_{OO'}\frac{T_s}{\text{s}}\int_{0}^{\infty}\frac{df}{\text{Hz}}\frac{1}{N_t}\sum_{i=1}^{N_t}\left(\tilde R_{OO'}^{B}(f,t_i)\tilde R_{OO'}^{G*}(f,t_i)+\text{cc.}\right)=\nonumber\\
&=\frac 1 2\sum_{OO'}\frac{T_s}{\text{s}}\int_{0}^{\infty}\frac{df}{\text{Hz}}\left(\tilde R_{OO'}^{B}(f)\tilde R_{OO'}^{G*}(f)+\text{cc.}\right)\;,\nonumber\\
\mathcal{F}_{\Omega_G\Omega_G}&=\sum_{OO'}\frac{T_s}{\text{s}}\int_{0}^{\infty}\frac{df}{\text{Hz}}\frac{1}{N_t}\sum_{i=1}^{N_t}\left|\tilde R_{OO'}^{G}(f,t_i)\right|^2=\sum_{OO'}\frac{T_s}{\text{s}}\int_{0}^{\infty}\frac{df}{\text{Hz}}\left|\tilde R_{OO'}^{G}(f)\right|^2\left(1+\frac{A^2}{2}\right)\;. 
\label{Fisher-f-t-conA}
\end{align}
From the above discussion, we then see that the elements of the Fisher matrix for the sub-optimal frequency-only analysis are also given by eqs.~\eqref{Fisher-f-t-conA}, if one simply sets $A = 0$ in the $\mathcal{F}_{\Omega_G\Omega_G}$ element. From the value of $A$ given above, we see that this term induces an increase of only $A^2/2\sim 1.7\%$ in this element in the time-frequency vs. the frequency-only analysis.
We now employ these Fisher matrix elements to plot the contour levels of the two analyses, and we show that we can reproduce the near coincidence of their results observed in figure \ref{fig:fisher_contours} for the LISA-self case. Specifically, let us consider the $3\sigma$ contours (although identical considerations can be done for all the other confidence levels), that we show in figure \ref{LISA3sigma-zoom}, to appreciate the difference between the two analyses. We note that the optimal analysis results in a slightly better determination of the parameter (smaller area of the $3\sigma$ region), which was not noticeable by eye with the scale shown in figure \ref{fig:fisher_contours}.
\begin{figure}[h!]
\centerline{
\includegraphics[width=0.7\textwidth,angle=0]{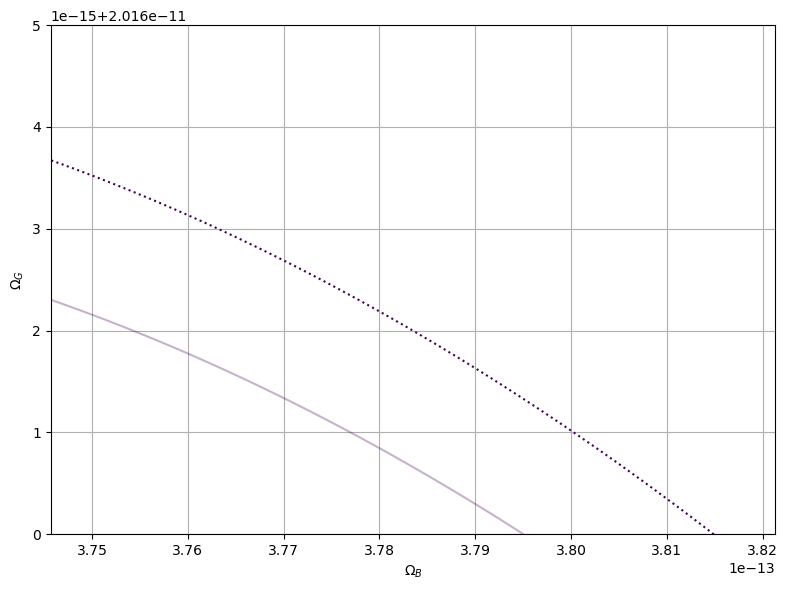}
}
\caption{A detail of figure \ref{fig:fisher_contours}, to visually appreciate the difference between the $3\sigma$ contours in the LISA-self analyses in the time-frequency domain (solid line) and in the sub-optimal frequency only case (dotted line). The central value of $\Omega_B$ on the horizontal axis corresponds to the fiducial value $\hat\Omega_B$.}
\label{LISA3sigma-zoom}
\end{figure}
From the Fisher matrix elements, the normalized posterior distribution of the two parameters $\Omega_B$ and $\Omega_G$, assuming a flat prior, is
\begin{align}
P(\Omega_B,\Omega_G)=\frac{1}{\sqrt{4\pi^2|(2\mathcal{F})^{-1}|}}e^{-\Theta^T\cdot\mathcal{F}\cdot\Theta}
\;\;,\;\; 
\mathcal{F}=\left(\begin{array}{cc}
\mathcal{F}_{\Omega_B\Omega_B}   &  \mathcal{F}_{\Omega_B\Omega_G} \\
\mathcal{F}_{\Omega_B\Omega_G}  & \mathcal{F}_{\Omega_G\Omega_G} 
\end{array}\right) \, , 
\end{align}
where $\Theta=(\Omega_B-\hat\Omega_B,\Omega_G-\hat\Omega_G)$, and we recall that $\hat\Omega_B = 3.78\times 10^{-13}$ and $\hat\Omega_G = 2\times 10^{-11}$ are the fiducial values used in the analysis. 
For each analysis, we fix $\Omega_B=\hat\Omega_B$, and we compute the value of $\Omega_G$ leading to the $3\sigma$ confidence level (that corresponds to the intersection of each curve in Figure \ref{LISA3sigma-zoom} with the vertical  $\Omega_B = 3.78\times 10^{-13}$ line). Denoting this value by $\tilde\Omega_G$, we have 
\begin{align}
(\tilde\Omega_G-\hat\Omega_G)^2\,\mathcal{F}_{\Omega_G\Omega_G}=3 \;\; \Rightarrow \;\; 
\tilde\Omega_G=\hat\Omega_G+\sqrt{\frac{3}{\mathcal{F}_{\Omega_G\Omega_G}}} \;. 
\end{align}
We recall that the Fisher matrix elements $\mathcal{F}_{\Omega_G\Omega_G}$ of the two analyses differ by the $1+\frac{A^2}{2}$ factor, leading to a discrepancy of $\tilde\Omega_G$ obtained in the two case quantified as 
\begin{align}
\Delta\tilde\Omega_G=\sqrt{\frac{3}{\mathcal{F}_{\Omega_G\Omega_G}}}-\sqrt{\frac{3}{\mathcal{F}_{\Omega_G\Omega_G}\left(1+\frac{A^2}{2}\right)}} = \frac{A^2}{4}\sqrt{\frac{3}{\mathcal{F}_{\Omega_G\Omega_G}}} + {\rm O } \left( A^4 \right) \;. 
\end{align}
As we already mentioned, $A = 0.185$. We numerically find $\frac{1}{\sqrt{\mathcal{F}_{\Omega_G\Omega_G}}}\simeq 7.6\times 10^{-14}$, therefore we have $\Delta\tilde\Omega_G\simeq 1.1\times 10^{-15}$, in excellent agreement with the distance between the two curves seen in figure \ref{LISA3sigma-zoom}.

\newpage
\bibliographystyle{apsrev}
\bibliography{paper-biblio}
	
\end{document}